\documentclass[12pt]{article}

\newcommand{\be}{\begin{eqnarray}}
\newcommand{\ee}{\end{eqnarray}}

\usepackage{amsfonts,amsmath}
\usepackage{latexsym}

\newcommand{\la}{\lambda}
\renewcommand{\a}{\alpha}
\newcommand{\G}{\Gamma}
\newcommand{\g}{\gamma}

\renewcommand{\d}{\delta}

\newcommand{\s}{\sigma}

\newcommand{\nn}{\nonumber}
\newcommand{\eps}{\epsilon}
\newcommand{\ve}{\varepsilon}
\newcommand{\cL}{{\cal{L}}}

\newcommand{\cD}{{\cal{D}}}

\newcommand{\cP}{{\cal{P}}}

\newcommand{\E}{E_{10}}
\newcommand{\KE}{K(E_{10})}
\newcommand{\KO}{E_{10}/K(E_{10})}

\newcommand{\lak}{{\mathfrak{k}}}

\newcommand{\soa}{SO(9)\times SO(9)}
\newcommand{\sob}{SO(9)\times SO(2)}
\newcommand{\bi}{\bar{\imath}}
\newcommand{\bj}{\bar{\jmath}\,}
\newcommand{\bk}{\bar{k}}
\newcommand{\bl}{\bar{l}}
\newcommand{\bal}{{\bar{\alpha}}}

\begin{document}
{\flushright AEI-2006-016\\[10mm]}

\begin{center}
{\bf \Large IIA and IIB spinors from $\KE$}\\[7mm]
Axel Kleinschmidt and Hermann Nicolai\\[5mm]
{\sl  Max-Planck-Institut f\"ur Gravitationsphysik\\
     Albert-Einstein-Institut \\
     M\"uhlenberg 1, D-14476 Potsdam, Germany} \\[7mm]
\begin{minipage}{12cm}\footnotesize
\textbf{Abstract:} We analyze the decomposition of recently
 constructed unfaithful spinor representations of $\KE$ under
 its $SO(9)\times SO(9)$, and $SO(9)\times SO(2)$ subgroups, respectively,
 where $\KE$ is the `maximal compact' subgroup of the hyperbolic Kac--Moody
 group $\E$. We show that under these decompositions, respectively,
 {\em one and the same} $\KE$ spinor gives rise to both the
 fermionic fields of IIA supergravity, {\em and} to the (chiral) fermionic
 fields of IIB supergravity. This result is thus the fermionic
 analogue of the decomposition of $\E$ under its $SO(9,9)$ and
 $SL(9)\times SL(2)$ subgroups, respectively, which yield the
 correct bosonic multiplets of (massive) IIA and IIB
 supergravity. The essentially unique Lagrangian for the
 supersymmetric $\KO$ $\s$-model therefore can also capture the
 dynamics of IIA and IIB including bosons and fermions in the known
 truncations. \\
\end{minipage}
\end{center}

\begin{section}{Introduction}

Recent work has established the existence of two unfaithful spinorial
representations of the infinite-dimensional,
`maximal compact' subgroup $\KE$ of the hyperbolic
Kac--Moody group $\E$, namely a `Dirac-spinor-type' representation
with 32 real
components \cite{dBHP05a}, and a `gravitino-type' representation with
320 real components~\cite{DKN,dBHP05b,DKNin}.\footnote{The corresponding
unfaithful representations for $K(E_9)$ had already been identified
and studied in \cite{NiSa05}.} In this letter, we analyze the decomposition
of these two representations under the maximal finite-dimensional subgroups
$SO(9)\times SO(9)$ and $SO(9)\times SO(2)$ of $\KE$, respectively, and
show that the `$\KE$-gravitino' $\bf{320}$ and the `$\KE$-Dirac-spinor'
$\bf{32}$ decompose and transform correctly as required
by the fermionic multiplets of IIA and IIB supergravity, respectively.
These consist, respectively, of two gravitini and dilatini, each pair
with both chiralities, for the IIA theory; and two gravitini of one
chirality, and two dilatini, of the opposite chirality, for IIB.
The decomposition of the `Dirac-spinor' representation $\bf{32}$
similarly yields the correct supersymmetry parameters of these theories.
We thus find that {\em one and the same} 320 component (and 32 component)
$\KE$ spinor thus gives rise to the fermions of both IIA {\em and} IIB
supergravity, depending on how one `slices' the infinite-dimensional group
$\KE$ under its finite-dimensional subgroups. This is our main result:
it extends previous ones on the emergence of the bosonic multiplets
of these theories from $\E$ under appropriate `level decompositions'
of $\E$ under its $A_9$, $D_9$ and $A_8\times A_1$ subgroups, respectively,
\cite{DaHeNi02,KlNi04a,KlNi04b}, and earlier results
on the embedding of these theories
into $E_{11}$ \cite{We01,SchnWe01,SchnWe02,KlSchnWe04,We04}.
The present results thus
strengthen the case for the (essentially unique) supersymmetric
$\KO$ $\s$-model proposed in \cite{DKN} as a candidate for a unification
of the maximally extended supergravity theories in ten and eleven
space-time dimensions into a single theory.

This paper has the following structure. In section~\ref{kesec},
we review, following \cite{DKN},
the definition and `low level' commutation relations of
$\KE$ viewed from its $SO(10)$ subgroup, and we
identify the $\soa$ and $\sob$ subgroups of $\KE$, relevant for
the IIA and IIB theories. In section~\ref{spindec}, we decompose
the unfaithful `Dirac-spinor' and `gravitino' representations of
$\KE$ under these subgroups. We also briefly discuss the relation
to type I supergravity and $DE_{10}$.
We end with some concluding remarks
in section~\ref{conc}.

\end{section}

\begin{section}{$SO(9)\!\times\! SO(9)$ and $SO(9)\!\times\! SO(2)$ subgroups}
\label{kesec}

The maximal compact group $\KE$ is defined as the subgroup of $\E$
whose Lie algebra is invariant under the Chevalley involution $\theta$
(see
\cite{Kac} for an introduction to the theory of Kac--Moody algebras).
In this section, we will study its distinguished subgroups $\soa$
and $\sob$, starting from the $SO(10)$ decomposition of $\KE$
at low levels presented in \cite{DKN}. These two subgroups together
generate all of $\KE$; there is no finite-dimensional R symmetry in
the present scheme that would accommodate the fermions of both IIA
and IIB supergravity.

\begin{subsection}{$\KE$ in terms of $SO(10)$}

In the approach of \cite{DKN}, the $\KE$ algebra was written in terms
of generators derived from the $SL(10)$ decomposition of $E_{10}$
\cite{DaHeNi02}. Up to $SL(10)$ level $\ell=3$, the generators
of the associated Lie algebra $\lak_{10}\equiv {\rm Lie}\, (\KE)$ are
defined by
\begin{align}\label{kso10}
J^{ab} &= K^a{}_b - K^b{}_a\,, &\quad J^{a_1a_2a_3}
  &= E^{a_1a_2a_3} - F_{a_1a_2a_3}\,,&\nn\\
J^{a_1\ldots a_6} &= E^{a_1\ldots a_6}- F_{a_1\ldots
  a_6}\,,&\quad
   J^{a_0|a_1\ldots a_8} &= E^{a_0|a_1\ldots a_8}-
  F_{a_0|a_1\ldots a_8}\,,&
\end{align}
with the $GL(10)$ generators $K^a{}_b$ and the basic (level $\pm 1$)
$\E$ generators $E^{abc}$ and $F_{abc} = -\theta(E^{abc}) \equiv
(E^{abc})^T$; thus, in terms
of those generators, the compact generators are generically
`anti-symmetric', i.e. of type $J= E-E^T \equiv E-F$. 
Henceforth, we shall refer to $J^{ab},\,J^{a_1a_2a_3}$,
$J^{a_1\ldots a_6}$, and $J^{a_0|a_1\ldots a_8}$ as being of `levels'
$\ell=0,1,2,3$, respectively, although this `level' is not a grading of
$\lak_{10}$~\footnote{Instead, we have a so-called `filtration'
$\big[\lak^{(\ell)},\,  \lak^{(\ell')}\big] \subset \lak^{(\ell+\ell')}
\oplus\lak^{(|\ell-\ell'|)}$. This is tantamount to introducing an
equivalence relation modulo `lower level' contributions, which again yields
a gradation over the natural numbers.}; in fact, $\lak_{10}$ is not
even a Kac--Moody algebra \cite{KN}. Up to $\ell=3$
(and neglecting higher level contributions), the $\lak_{10}$ commutation
relations are given by \cite{DKN,dBHP05b}~\footnote{Neglecting the
non-trace part of ${J}^{a_0|a_1\ldots a_8}$, the corresponding
commutators for $K(E_{11})$ were already computed in \cite{We03}.}
\be\label{KE10}
\left[J^{ab}, J^{cd}\right] &=&
  \d^{bc}J^{ad}+\d^{ad}J^{bc}-\d^{ac}J^{bd}-\d^{bd}J^{ac}  \equiv
  4 \d^{bc}J^{ad}\nn\\
\left[J^{a_1a_2a_3}, J^{b_1b_2b_3}\right] &=& J^{a_1a_2a_3b_1b_2b_3} -
  18 \d^{a_1b_1}\d^{a_2b_2}J^{a_3b_3}\nn\\
\left[ J^{a_1a_2a_3}, J^{b_1\ldots b_6}\right] &=&
  J^{[a_1|a_2a_3]b_1\ldots
    b_6}- 5!\,\d^{a_1b_1}\d^{a_2b_2}\d^{a_3b_3}J^{b_4b_5b_6}\nn\\
\left[J^{a_1\ldots a_6}, J^{b_1\ldots b_6}\right] &=& -6\cdot
   6!\,\d^{a_1b_1}\cdots \d^{a_5b_5} J^{a_6b_6}+\ldots\nn\\
\left[J^{a_1a_2a_3}, J^{b_0|b_1\ldots b_8}\right] &=& -336\,\left(
  \d^{b_0b_1b_2}_{a_1a_2a_3}J^{b_3\ldots b_8} -
  \d^{b_1b_2b_3}_{a_1a_2a_3} J^{b_4\ldots b_8b_0} \right)+\ldots\nn\\
\left[J^{a_1\ldots a_6}, J^{b_0|b_1\ldots b_8}\right] &=&
  - 8!\,\left(\d^{b_0b_1\ldots b_5}_{a_1\ldots a_6} J^{b_6b_7b_8} -
  \d^{b_1\ldots b_6}_{a_1\ldots a_6} J^{b_7b_8b_0}\right)+\ldots\nn\\
\left[J^{a_0|a_1\ldots a_8}, J^{b_0|b_1\ldots b_8}\right]
  &=& -8\cdot 8!\,\left(\d^{a_1\ldots a_8}_{b_1\ldots b_8} J^{a_0b_0} -
  \d^{a_1\ldots a_8}_{b_0b_1\ldots b_7} J^{a_0b_8} - \d^{a_0a_1\ldots
  a_7}_{b_1\ldots b_8} J^{a_8b_0}\right.\nn\\
&&\quad \left.+ 8 \,\d^{a_0}_{b_0} \d^{a_1\ldots a_7}_{b_1\ldots b_7}
  J^{a_8b_8} +7 \d^{a_1}_{b_0} \d^{a_0a_2\ldots a_7}_{b_1\ldots b_7}
  J^{a_8b_8}\right)+\ldots
\ee
Here, all indices $a,b,\dots = 1,\dots,10$ are to be regarded as (`flat')
$SO(10)$ indices. As in \cite{DKN} we make use of a shorthand notation
in (\ref{KE10}), where the terms on the r.h.s. are to be
anti-symmetrized (with weight one) according to the anti-symmetries on
the l.h.s., as explicitly written out for the $SO(10)$ generators
$J^{ab}$ in the first line. As also explained there, the
generator $J^{a_0|a_1\ldots a_8}$ decomposes into two irreducible
components under $SO(10)$, the trace and a non-trivial mixed
(Young tableau) representation.\footnote{Whereas $J^{a_0|a_1\ldots a_8}$
is irreducible as a representation of $SL(10)$.}
The $SO(10)$ generators rotate the higher level generators in the
standard way as tensor representations.

\end{subsection}

\begin{subsection}{$SO(9)\times SO(9)$ subgroup}

In ref.~\cite{KlNi04a}, the $E_{10}$ group was analyzed under its regular
$SO(9,9)$ subgroup. Similarly, one can study $\KE$ under its
$K(SO(9,9))= \soa$ subgroup. The corresponding generators
of $\soa$ can be written in terms of the $SO(10)$ generators of
(\ref{kso10}) as
\be\label{kso9so9}
X^{ij} &:=& \frac12\left(J^{(0) ij} + J^{(1) ij10}\right) ,\nn\\
X^{\bi\bj} &:=& \frac12\left(J^{(0) ij} - J^{(1) ij10}\right) \qquad
  \mbox{for $i,j=1,\dots,9$}.
\ee
We use the notation of \cite{KlNi04a} with barred indices for the
second $SO(9)$ in order to distinguish the two factors.
It is worthwhile to note that this definition uses both `level zero'
and `level one' generators in terms of (\ref{kso10}). We have
indicated the $A_9$ `level' explicitly in parentheses in
(\ref{kso9so9}). The form of
these generators can be found by tracing back the definitions of
$X^{ij}$ and $X^{\bi\bj}$ in terms of the Chevalley generators of
$\E$ given in \cite{KlNi04a} and then re-expressing them in terms
of the $SO(10)$ tensors (\ref{kso10}).

From  (\ref{KE10}) it is straight-forward to check that the
generators (\ref{kso9so9}) satisfy the $SO(9)\times SO(9)$
commutation relations
\be
\left[X^{ij}, X^{kl}\right] &=& \d^{jk} X^{il} + \d^{il} X^{jk} -\d^{jl}
X^{ik}-\d^{ik} X^{jl},\nn\\
\left[X^{\bi\bj}, X^{\bk\bl}\right] &=& \d^{\bj\bk} X^{\bi\bl}
  + \d^{\bi\bl} X^{\bj\bk} -\d^{\bj\bl}X^{\bi\bk}
  -\d^{\bi\bk} X^{\bj\bl},\nn\\
\left[X^{ij}, X^{\bk\bl}\right] &=& 0 \;\; ,
\ee
Evidently, no `higher level' ($\ell\geq 2$)
generators of $\KE$ are excited in these commutators.

\end{subsection}

\begin{subsection}{$\sob$ subgroup}

The analysis of \cite{KlNi04b} started from an $SL(9)\times SL(2)$
decomposition of $E_{10}$ where the $SL(2)$ was identified with the
$SL(2)$ symmetry of type IIB supergravity after establishing a
dynamical correspondence. At the level of compact subgroups the
decomposition is $\sob\equiv K(SL(9)\times SL(2))\subset \KE$ and the
generators are defined as
\be\label{kso92}
R^{rs} &:=& J^{(0)rs} \;\; , \;\;\; R^{r9} := -R^{9r}:=J^{(1) r9\,10}
\;\; , \;\;\; \qquad \mbox{for $r,s=1,\dots,8$}\nn\\
R &:=& J^{(0)9\,10} .
\ee
The nine $SO(9)$ indices had to be split into $(8+1)$ since eight
directions are in common with an $SO(8)\subset SO(10)$ but one
direction is different. This is in line with standard views on
T-duality \cite{DHS,DLP,AdWLN,Ha00}. The expressions (\ref{kso92}) can
again be deduced by going via the explicit relation to the standard
Chevalley basis of $\E$.

From (\ref{KE10}) one can show that the generators (\ref{kso92})
satisfy the $SO(9)\times SO(2)$ relations
\be
\left[R^{ij}, R^{kl}\right] = 4\d^{jk} R^{il}\quad,\quad\quad
\left[R, R^{ij}\right] = 0
\ee
for $i,j,k,l=1,\ldots,9$ (again with anti-symmetrizations understood).
In particular, $R$ commutes with all $SO(9)$ generators, as required.
In the supergravity context the compact $SO(2)$ is usually
referred to as $U(1)$ and we will use both terms interchangeably.

\end{subsection}

\end{section}

\begin{section}{Unfaithful $\KE$ spinor representations}
\label{spindec}

In \cite{dBHP05a,DKN,dBHP05b} two unfaithful representations of $\KE$
were defined. These consist of 32 and 320 real components, respectively.
In terms of the $SO(10)$ generators (\ref{kso10}), the `Dirac spinor'
representation $\bf{32}$, denoted by $\eps$, transforms as \cite{dBHP05a}
\be\label{dstrm}
J^{(0)ab}\cdot \eps = \frac12 \G^{ab}\eps\quad,\quad\quad
J^{(1)abc}\cdot \eps = \frac12 \G^{abc}\eps,
\ee
The `vector-spinor' representation $\bf{320}$, denoted by $\psi^a$, transforms
as \cite{DKN,dBHP05b}
\be\label{vstrm}
J^{(0)ab}\cdot\psi^{c} &=& \frac12\G^{ab}\psi^c + 2 \d^{c[a}\psi^{b]},\nn\\
J^{(1)abc}\cdot\psi^{d} &=& \frac12 \G^{abc}\psi^d
    + 4\d^{d[a}\G^b\psi^{c]} - \G^{d[ab}\psi^{c]}.
\ee
Here, $\G^{a}$ ($a=1,\ldots,10$) are the real $(32 \times 32)$ spatial
$SO(1,10)$ $\G$-matrices
of eleven-dimensional supergravity \cite{CJS}, which we use in the basis
defined in~\cite{KlNi04a}.
In both cases we have given the transformations only up to
$SO(10)$ `level one' since this suffices to characterize the
consistent unfaithful $\KE$ representation \cite{DKNin}.\footnote{The
  transformation rules up to `level three' are known and can be found in
  \cite{dBHP05a,DKN,dBHP05b}.} Furthermore, the $\soa$ and $\sob$
subgroup generators (\ref{kso9so9}) and (\ref{kso92}) only require the
$SO(10)$ levels zero and one. The transformation rules
(\ref{dstrm}) and (\ref{vstrm}) were derived in
\cite{DKN,dBHP05b}
by demanding a dynamical correspondence between a $\KE$ covariant
spinor equation and the gravitino equation of motion of $D=11$
supergravity.

In the following we will decompose the Dirac- and vector-spinor under
the $\soa$ and $\sob$ subgroups of $\KE$. For this it will be
important that the irreducible Dirac-spinor of $SO(9)$ has 16 real
components and that in our basis the $(32\times 32)$ matrix $\G^{10}$
is of the block diagonal form
\be
\G^{10}=\left(\begin{array}{cc}{\bf 1}_{16}&0\\0&-{\bf
    1}_{16}\end{array}\right),
\ee
The combinations
\be\label{proj}
P_\pm = \frac12\big[1\pm\G^{10}\big]
\ee
are the standard orthogonal projectors onto two $16$-component subspinors
of the $32$ component Majorana spinor of $SO(1,10)$; viewed from $D=10$,
they become the two spinors of IIA supergravity of opposite chirality.

\begin{subsection}{$\soa$ decomposition}

\begin{subsubsection}{Dirac-spinor}

We diagonalize the action of (\ref{kso9so9}) on the $\KE$
Dirac-spinor $\eps$ by defining
\be
\ve_\pm = P_\pm\eps
\ee
using the projectors (\ref{proj}).
We immediately check the action of the first $SO(9)$ factor $\soa$
from (\ref{dstrm}) as
\be
X^{ij}\cdot \ve_\pm =
  \frac14 P_\pm\G^{ij}(1+\G^{10})\eps
  = \frac12 \G^{ij} P_\pm P_+\eps,
\ee
such that $\ve_+$ transforms as a Majorana spinor under this
$SO(9)$, and $\ve_-$ transforms trivially.~\footnote{The matrices
$\G^{ij}P_\pm$ can be identified with the  $(16\times 16)$ gamma matrices
of $SO(9)$.} Under the second $SO(9)$ these
properties are obviously interchanged, and therefore we deduce the
following decomposition of the $\KE$ Dirac-spinor under $\soa$
\be\label{dsbrancha}
{\bf 32} \quad\longrightarrow\quad ({\bf 1},{\bf
    16})\oplus({\bf 16},{\bf   1})
\ee
Under the diagonal $SO(9)_{\rm diag}$ these two spinors become
two spinors of opposite handedness~\footnote{This terminology
obviously refers to the chirality w.r.t. to the space-time Lorentz
group $SO(1,9)$, from which these spinors originate. Although $SO(9)$
by itself has no chiral representations, we will nevertheless make
occasional use of it.}.

\end{subsubsection}

\begin{subsubsection}{Vector-spinor}

In order to find the irreducible components of the $\KE$
vector-spinor (or gravitino) $\psi_a$ in terms of $\soa$
representations we have to redefine the fermionic
components according to \cite{KlNi04a} as
\be\label{fermredef}
\tilde{\psi}_k &=& \psi_k +\frac12\G_k\G^{10}\psi_{10},\nn\\
\tilde{\psi}_{10} &=& -\frac32\psi_{10} -\G_{10}\G^k\psi_k.
\ee
For example, acting with $X^{ij}$ on $P_-\tilde{\psi}_{10}$, using
the formulas (\ref{vstrm}) leads to
(after some computation)\footnote{Of course, (\ref{dstrm}) does not fix
  the normalization of $\tilde{\psi}_{10}$ but only the relative
  coefficients. The normalization was fixed in \cite{KlNi04a} by
  demanding a canonical form of the resulting kinetic term.}
\be\label{dstrmso9so9}
X^{ij}\cdot P_-\tilde{\psi}_{10}
 = \frac12 \G^{ij}P_-\tilde{\psi}_{10},
\ee
so that the $K(E_{10})$ action (\ref{vstrm}) expressed in terms of
$SO(9)\times SO(9)$ via (\ref{kso9so9}) is equivalent to the
projection on one chiral component which transforms in the usual
$SO(9)$ spinor representation. In other words,
$P_-\tilde{\psi}_{10}$ transforms in the
${\bf 16}$ representation under the first $SO(9)$ of
$SO(9)\times SO(9)$ and trivially under the second $SO(9)$ (the
projectors are orthogonal). The opposite behaviour is
deduced for $P_+\tilde{\psi}_{10}$, so that
the $\tilde{\psi}_{10}$ part of the ${\bf 320}$ of $K(E_{10})$
gives rise to the $({\bf 16},{\bf 1})\oplus({\bf 1},{\bf 16})$ under
$\soa$. A similar calculation for the $\tilde{\psi}_k$ component gives
\be
X^{ij}\cdot\tilde{\psi}^{k} &=& \frac12\G^{ij}P_+\tilde{\psi}^k
   +2\d^{k[i}P_-\tilde{\psi}^{j]},\nn\\
X^{\bi\bj}\cdot\tilde{\psi}^{k} &=& \frac12\G^{ij}P_-\tilde{\psi}^k
   +2\d^{k[i}P_+\tilde{\psi}^{j]}.
\ee
The two different projectors appearing in this computation imply
that the $P_+\tilde{\psi}^k$ part transforms in the spinor ${\bf 16}$
of the first $SO(9)$, and the
vector ${\bf 9}$ of the second $SO(9)$, and oppositely for the other
chiral projection $P_-\tilde{\psi}^k$.

Together with (\ref{dstrm}) we therefore deduce the expected total
result
\be\label{vsbranch}
{\bf 320} \quad\longrightarrow\quad
  ({\bf 9},{\bf 16})\oplus({\bf 1},{\bf  16}) \oplus
  ({\bf 16},{\bf 9})\oplus({\bf 16},{\bf  1}).
\ee
This is precisely the representation used in \cite{KlNi04a} for which
also a partial dynamical check was carried out there, analogous to the
one performed in \cite{DaHeNi02}. The representations on the
r.h.s. are to be interpreted as the two gravitini and the two
dilatini of (massive) IIA supergravity.
We emphasize again the chirality-symmetric
nature of these spinors, which here manifests itself in the symmetry
under interchange of the two $SO(9)$ groups.

\end{subsubsection}

\end{subsection}

\begin{subsection}{$\sob$ decomposition}

For the analysis of the subgroup $\sob$, we first define the matrix
\be
\G^* := \G^9 \G^{10}.
\ee
Obviously, $(\G^*)^2 = -{\bf 1}$, whence $\G^*$ can be regarded as
an imaginary unit. This will turn out to be a main difference
with the $\soa$ decomposition of the previous section: unlike $\G^{10}$
(which squares to $+{\bf 1}$), we cannot use $\G^*$ to define projectors
unless we complexify the representation. Two useful relations are
\be
P_\pm\G^* = \G^*P_\mp,\quad\quad P_\pm \G^{10} = \pm P_\pm\nn.
\ee

\begin{subsubsection}{Dirac-spinor}

For the $\KE$ Dirac-spinor we define new components via
\be
\ve_1 := P_- \eps \quad , \quad
\ve_2 := P_- \G^* \eps = \G^* P_+ \eps\quad.
\ee
The use of the projector $P_-$ here makes explicit that we are really dealing
with two 16-component objects, from which the original spinor can be
reconstructed via
\be
\eps = \ve_1 - \G^* \ve_2
\ee
Equivalently, we could work with the complex 16-component spinor
$\ve_1 -i\ve_2$, replacing the matrix $\G^*$ by the imaginary
unit.

Acting with an $SO(2)$ rotation $R$ from (\ref{kso92}) on this spinor
we obtain
\be
R\cdot \ve_1 = + \frac12 \ve_2 \quad , \qquad
R\cdot \ve_2 = - \frac12 \ve_1\quad.
\ee
The pair $(\ve_1,\ve_2)$ therefore transforms as a doublet under
$SO(2)$: the complex spinor $\ve_1 \pm i\ve_2$ carries
$U(1)$ charge $\mp\frac12$.

Under $SO(9)$ the components $\ve_1$ and $\ve_2$ both transform as
(suppressing the $SO(2)$ indices)
\be\label{dstrmso9}
R^{rs}\cdot \ve = \frac12 \G^{rs} \ve \quad,\quad\quad
R^{r9}\cdot \ve = +\frac12 \G^r\G^* \ve \;\;.
\ee
This is the correct transformation of an $SO(9)$ Dirac-spinor, as it
does not mix $\ve_1$ and $\ve_2$ (both $\G^{rs}$ and $\G^r\G^*$
commute with $P_\pm$ and $\G^*$). Due to the presence of
the projector $P_-$ in the definition of $\ve_1$ and $\ve_2$
the action of the $(32\times 32)$ $\G$-matrices of $SO(1,10)$
can be seen as the action of $(16\times 16)$ $\g$-matrices of $SO(9)$.
The present realization of the $SO(9)$ Clifford algebra is in terms of
$(32\times 32)$ matrices $\left\{\G^{rs},\G^r\G^*\right\}$.
We conclude that the {\bf 32} of $\KE$ decomposes under
the $\sob$ subgroup into an $SO(2)$ doublet of $SO(9)$ spinors:
\be\label{dsbranchb}
{\bf 32} \quad\longrightarrow\quad ({\bf 16},{\bf 2}).
\ee

\end{subsubsection}

\begin{subsubsection}{Vector-spinor}

The $\KE$ vector-spinor $\psi_a$ gives rise to two kinds of spinors,
namely the dilatini
\be\label{newferm1}
\la_1 &:=& P_-\left(\psi_9 - \G^* \psi_{10}\right),
\nn\\
\la_2 &:=& P_-\left(\G^*\psi_9 + \psi_{10}\right),
\ee
and the gravitini
\be\label{newferm2}
\chi_1^9 &:=& P_- \left(\psi_9 + \G^* \psi_{10}\right), \nn\\
\chi_2^9 &:=& P_- \left(-\G^* \psi_9 +\psi_{10}\right)
    = -P_-\G^*\left(\psi_9+\G^* \psi_{10}\right), \nn\\
\chi_1^r &:=& P_- \left(-\frac43\psi^r -\frac13 \G^r\G^9\psi_9
  -\frac13 \G^r \G^{10}\psi_{10}\right), \nn\\
\chi_2^r &:=& P_- \G^*
  \left(-\frac43\psi^r -\frac13 \G^r\G^9\psi_9
  -\frac13 \G^r \G^{10}\psi_{10}\right).
\ee
Again, the lower index $(1,2)$ is the $SO(2)$ index, while the
upper index (with $r=1,\ldots,8$ as in (\ref{kso92})) is an
$SO(9)$ vector index. The linear combinations appearing in the
above equation were arrived at by demanding proper behavior under
$SO(9)$ transformations, to wit, by requiring that $\chi^9$ and
$\chi^r$ combine into a vector-spinor $\chi^i$ ($i=1,\ldots,9$) of
$SO(9)$, starting from (\ref{dstrm}) and (\ref{vstrm}), see below.
As before, one can also combine
these spinors into a complex dilatino
$\la\equiv\la_1 \pm i \la_2$ and a complex gravitino
$\chi^i\equiv\chi^i_1\pm i\chi_2^i$.

First we determine the $SO(2)$ properties of the new fermions (\ref{newferm1})
and (\ref{newferm2}). A straightforward calculation using (\ref{kso92})
and (\ref{vstrm}) together with (\ref{newferm1}) shows that
\be
R\cdot \la_1 = + \frac32 \la_2 \quad , \qquad
R\cdot \la_2 = - \frac32 \la_1 \quad,
\ee
whence the complex dilatino $\la_1 \pm i\la_2$ carries $U(1)$ charge
$\mp\frac32$.
The pair of gravitini turns out to have the same $SO(2)$ charge as
the Dirac-spinor representation (in agreement with the fact that the gravitino
and the supersymmetry transformation parameter should transform in the
same $SO(2)$ representation). Using  (\ref{kso92}) and (\ref{vstrm})
again, we derive
\be
R\cdot \chi^i_1 = + \frac12 \chi^i_2 \quad , \qquad
R\cdot \chi^i_2 = - \frac12 \chi^i_1 \quad .
\ee

Under the $SO(9)$ part $\sob$ one finds after some computation that
\be
R^{rs}\cdot \la =  \frac12 \G^{rs}\la \quad , \qquad
R^{r9}\cdot \la = -\frac12 \G^{r}\G^* \la.
\ee
for both $\la_1$ and $\la_2$. This is the transformation of an $SO(9)$
spinor where now the representation is in terms of $(32\times 32)$ matrices
$\left\{\G^{rs},-\G^r\G^*\right\}$, projected onto $(16\times 16)$ matrices
by the projector $P_-$ in the redefinition of the fermions.
Notice that this Clifford algebra now differs from
formula (\ref{dstrmso9}) by the sign in front of $\G^r\G^*$. Although
the representations are equivalent over $SO(9)$ (for which there is no
chirality), we take this difference of sign as a
manifestation of the opposite chiralities of these two spinors in
IIB supergravity w.r.t. the Lorentz group $SO(1,9)$ in ten dimensions
\cite{Schwarz:1983wa,Schwarz:1983qr}.
For the gravitino components (\ref{newferm2}) one finds similarly,
using (\ref{vstrm}) and the redefinitions (\ref{newferm2}),
\be
R^{rs}\cdot \chi^{k} &=&  \frac12 \G^{rs}\chi^{k} +
    2\d^{k[r}\chi^{s]}\,,\nn\\
R^{r9}\cdot \chi^k &=& +\frac12 \G^{r} \G^* \chi^k +
    2\d^{k[r}\chi^{9]} \,.
\ee
where $k=1,\dots,9$, and where we have suppressed the $SO(2)$ indices.
This is the correct transformation of a vector-spinor (with
$\G$-trace) under $SO(9)$. Therefore, we conclude that the $\KE$
vector-spinor decomposes as
\be\label{vsbranchb}
{\bf 320} \quad\longrightarrow\quad
  (\overline{{\bf 16}},{\bf 2})\oplus({\bf 144},{\bf  2})
\ee
under $\sob$, giving the dilatino and gravitino doublets (where the
bar over the first $\bf{16}$ is meant to indicate opposite chirality
w.r.t. $SO(1,9)$).\footnote{In the ${\bf 144}$ of $SO(9)$  a
$\g$-trace can be separated (whence the representation is reducible),
but we prefer to write the ${\bf 144}$ as one representation since
it corresponds to the physical gravitino.}
In summary, we have obtained perfect agreement between the IIB
supergravity assignments of the fermionic fields and the $SO(2)$
(or $U(1)$) charges and $SO(9)$ assignments as they emerge from
$\KE$!

\end{subsubsection}

\end{subsection}

\begin{subsection}{Embedding $K(DE_{10})\subset\KE$}

The truncation to pure type I theory in ten
space-time dimensions is conjectured to be associated with the
algebra $D\E$ in the hyperbolic KMA setting \cite{Ju85}
(for very-extended algebras one expects $DE_{11}$ \cite{SchnWe04}). It
was shown in \cite{KlNi04a} that $D\E$ is a proper subgroup of
$\E$. Moreover, it was shown there that in the $SO(9,9)$ decomposition of
$D\E$ only {\em tensorial} representations appear, whereas for the
decomposition of $\E$ one also finds $SO(9,9)$ spinor representations
(for example associated with the RR fields). Under the compact $\soa$
of $SO(9,9)$, an $SO(9,9)$ Dirac-spinor decomposes into the tensor
product of two $SO(9)$ spinors associated with the two $SO(9)$ factors
of $\soa$. Such a generator was denoted $E_{\a\bal}$ in \cite{KlNi04a}
and evidently can be used to transform an $\soa$ representation of the
type  $({\bf 1},{\bf 16})$ into $({\bf 16},{\bf 1})$, i.e. changing
the `chirality' under $\soa$. Since generators like $E_{\a\bal}$ do not
exist in $D\E$ this can never happen and we are led to conclude that
there are two inequivalent 16-dimensional Dirac-spinor representations
of $K(D\E)$ which we denote by ${\bf 16}$ and $\overline{\bf
  16}$. They are distinguished by their decomposition under $\soa$
via
\be
{\bf 16} \quad \longrightarrow \quad ({\bf 16},{\bf 1}),\quad\quad
\overline{\bf 16} \quad \longrightarrow \quad ({\bf 1},{\bf 16}).
\ee
We also find that under $K(D\E)\subset K(\E)$ the $\KE$ Dirac-spinor
decomposes as
\be
{\bf 32} \quad\longrightarrow\quad {\bf 16}\oplus\overline{\bf 16}.
\ee
Similarly, the decomposition of the $\KE$ vector-spinor is expected to
be
\be
{\bf 320} \quad\longrightarrow\quad {\bf 160}\oplus\overline{\bf 160}.
\ee
In this sense, and because it gives rise to $D=10$ spinors of a
given chirality, $K(DE_{10})$ can be viewed as a `chiral half'
of $\KE$.

\end{subsection}

\begin{subsection}{Unfaithful spinor representations of $K(E_n)$}

Our results can be extended to unfaithful spinor representations 
of $K(E_n)$ for any $n\ge 9$, most notably
$K(E_9)$ \cite{NiSa05} and $K(E_{11})$ \cite{We03}. The form of the
transformation rules (\ref{dstrm}) and (\ref{vstrm}) imply that they
define consistent unfaithful representation for the maximal compact
subgroup $K(E_n)$ of $E_n$ for $n\ge 9$.\footnote{For $n<9$, 
it is straightforward to check that the relevant maximal compact 
subgroups $K(E_8)\equiv Spin(16)/Z_2$, {\it etc.}, are faithfully 
generated by the matrices $\G^{ab}, \G^{abc}, \dots$} These representations 
are written in terms of the $SO(n)$ subgroups of $K(E_n)$, but one could
also introduce the flat metric of $SO(n-p,p)$ (with corresponding real
gamma matrices) in (\ref{vstrm}), in particular, $SO(1,10)$ for
$K(E_{11})$ by using the so-called temporal involution
\cite{EnHo04a,KlWe04}.\footnote{See \cite{Ke04} for an analysis of
  the orbits of non-Euclidean signatures under the $E_{11}$ Weyl
  group.} The reason that (\ref{dstrm}) and (\ref{vstrm}) define
unfaithful representations for any $K(E_n)$ ($n\ge 9$) is that the
necessary consistency conditions do not involve traces (and so are
independent of $n$) and take the same form for all $n\ge 9$.

The dimension of the Dirac-spinor representation of $K(E_n)$
is given by the dimension of the real spinor representation of $SO(n-p,p)$,
i.e. $16$ for $(n,p)=(9,0)$ and $32$ for $(n,p)=(11,1)$. The na\"ive
dimension of the vector-spinor is then $= 144$ for $K(E_9)$ and 
$= 352$ for $K(E_{11})$ (with temporal involution). However, the
resulting vector-spinor of $K(E_9)$ is reducible since one can
construct the gamma trace of the vector-spinor in a $K(E_9)$ invariant
fashion. Therefore the irreducible vector-spinor of $K(E_9)$ has
dimension $128$, as already established in \cite{NiSa05}. For all other 
$n>9$, the gamma matrices are not invariant objects under $K(E_n)$.
For $K(E_{11})$, the vector-spinor has dimension $352$, and we have 
checked that under the IIA and IIB decompositions of $K(E_{11})$
this ${\bf 352}$ reduces to the correct $SO(1,9)$ covariant 
vectorlike and chiral spinors of the IIA and IIB theories,
respectively.\footnote{The inner products on these
  vector-spinor representations are, however, not given by the same form
  as for $K(E_{10})$, as the invariance of the latter requires 
  ten dimensions \cite{DKN}.}  However, these considerations concern
the representations, and are therefore purely kinematical. We have not
investigated the form of the $K(E_{11})$ covariant spinor equations
corresponding to the $D=11$ gravitino variation and equation of
motion.

\end{subsection}

\end{section}

\begin{section}{Outlook}
\label{conc}

In this paper, we have demonstrated that, at the {\em kinematical} level,
the unfaithful ${\bf 32}$ and ${\bf 320}$ spinor representations of
$\KE$ decompose into the correct fermionic representations under
the subalgebras relevant for the IIA and IIB analysis of $\E$. At
the {\em dynamical} level, partial checks concerning the $\soa$
decomposed fermionic sector were carried out already in \cite{KlNi04a}.
Using the formulation of \cite{DKN}, the dynamical system
with manifest full local $\KE$ invariance (and global $\E$
invariance) is correctly described by the Lagrangian
\be\label{susyl}
\cL = \frac1{2n} \langle \cP | \cP \rangle  -i (\chi|\cD\chi)_{\rm{vs}}.
\ee
This $\KO$ $\s$-model describes the motion of a massless spinning
particle on the $\KO$
coset space, where we take the `spin'
in the ${\bf 320}$ unfaithful vector-spinor
representation of $\KE$. Here, $\cP$ is the $\KE$ covariant bosonic
velocity and $\chi$ the matter fermion with $\cD$ being the $\KE$ covariant
derivative. As shown in \cite{DKN} (see also \cite{dBHP05b}), the Lagrangian
(\ref{susyl}) correctly
reproduces both the fermionic and bosonic equations of motion of
eleven-dimensional supergravity at linearized fermion order and with
the appropriate truncations on the supergravity side, i.e. neglecting
second and higher order spatial gradients for the bosonic fields,
and all spatial gradients of the fermionic fields. As shown in 
\cite{KlNi04a}, the action of IIA supergravity reduces to a $\KE$-invariant
action of the type (\ref{susyl}). Given the kinematical
`versatility' of $\KE$ and these results for $D=11$ (massive) IIA
supergravity we expect that (\ref{susyl}) will also
describe the fermionic equations of motion of IIB supergravity (in
the same truncation). Moreover, the $\KE$-invariant model
(\ref{susyl}) would exhibit also a dynamical versatility!

In light of the gradient hypothesis of \cite{DaHeNi02},
it is also interesting to note that the {\em space-time dimension} of
the corresponding theory depends upon which subgroup of $\E$ (or $\KE$)
is selected to perform the level decomposition: $D=11$ for $A_9\equiv SL(10)$,
and $D=10$ for $D_9\equiv SO(9,9)$ and
$A_8\times A_1 \equiv SL(9)\times SL(2)$.
In this sense, the dimension of space-time
is no longer a fundamental {\it datum} of this theory, but an emergent
phenomenon. Let us also note that a different proposal for the emergence
of space-time and its dimension
within the framework of $E_{11}$ was already made in
\cite{We03,KlWe04} where, however, space-time is implemented in terms
of an $E_{11}$ representation rather than within the algebra itself.

To be sure, we envisage that ultimately the unfaithful,
finite-dimensional representations of $\KE$ studied here will be
replaced by faithful representations (probably by taking tensor
products with the coset or the compact subgroup itself), replacing
the 320 time-dependent gravitino components $\psi_a$ by a infinite
tower of components characterizing the spatial dependence of the
fermions in line with the gradient conjecture of
\cite{DaHeNi02}.\\

\end{section}

{\bf Acknowledgements}\\
The authors are grateful to Stefan Theisen for discussions.


\baselineskip14pt

\begin{thebibliography}{20}

\bibitem{dBHP05a} S.~de Buyl, M.~Henneaux and L.~Paulot, {\sl Hidden
  symmetries and Dirac fermions}, Class.\ Quant.\ Grav.\  {\bf 22}
  (2005) 3595, {\tt hep-th/0506009}

\bibitem{DKN} T.~Damour, A.~Kleinschmidt and H.~Nicolai, {\sl Hidden
  symmetries and the fermionic sector of eleven-dimensional
  supergravity}, Phys. Lett. {\bf B 634} (2006) 319,
  {\tt hep-th/0512163}

\bibitem{dBHP05b} S.~de Buyl, M.~Henneaux and L.~Paulot, {\sl Extended
  $E_8$ invariance of 11-dimensional supergravity}, {\tt
  hep-th/0512292}

\bibitem{DKNin} T.~Damour, A.~Kleinschmidt and H.~Nicolai, in
  preparation.

\bibitem{NiSa05} H.~Nicolai and H.~Samtleben, {\sl On $K(E_9)$},
  Q.\ J.\ Pure Appl.\ Math.\  {\bf 1} (2005) 180, {\tt hep-th/0407055}

\bibitem{DaHeNi02} T.~Damour, M.~Henneaux and H.~Nicolai, {\sl
    $E_{10}$ and a "small tension expansion" of M-theory}, Phys.
    Rev. Lett. {\bf 89} (2002) 221601, {\tt hep-th/0207267}

\bibitem{KlNi04a} A.~Kleinschmidt and H.~Nicolai, {\sl $E_{10}$ and
  $SO(9,9)$ invariant supergravity}, JHEP {\bf 0407} (2004) 041, {\tt
  hep-th/0407101}

\bibitem{KlNi04b}  A.~Kleinschmidt and H.~Nicolai, {\sl IIB
  supergravity and $E_{10}$}, Phys.\ Lett.\ B {\bf 606} (2005) 391,
  {\tt hep-th/0411225}

\bibitem{We01} P.~C.~West, {\sl $E_{11}$ and M theory}, Class.
    Quant. Grav. {\bf 18} (2001) 4443--4460, {\tt hep-th/0104081}

\bibitem{SchnWe01} I.~Schnakenburg and P.~C.~West, {\sl Kac-Moody
  symmetries of IIB supergravity}, Phys.\ Lett.\ B {\bf 517} (2001)
  421, {\tt hep-th/0107181}

\bibitem{SchnWe02} I.~Schnakenburg and P.~C.~West, {\sl Massive IIA
  supergravity as a non-linear realisation}, Phys.\ Lett.\ B {\bf 540}
  (2002) 137, {\tt hep-th/0204207}

\bibitem{KlSchnWe04} A.~Kleinschmidt, I.~Schnakenburg and P.~West,
    {\sl Very extended Kac--Moody algebras and their
    interpretation at low levels}, Class. Quant. Grav. {\bf 21}
    (2004) 2493--2525, {\tt hep-th/0309198}

\bibitem{We04} P.C.~West, {\sl The IIA, IIB and eleven-dimensional theory
 and their common $E_{11}$ origin}, {\tt hep-th/0402140}

\bibitem{Kac} V.~Kac, {\sl Infinite dimensional Lie algebras},
  Cambridge University Press (Cambridge, 1990)

\bibitem{KN} A.~Kleinschmidt and H.~Nicolai, {\sl Gradient representations
  and affine structures in $AE_n$}, Class. Quant. Grav. {\bf 22} (2005)
  4457, {\tt hep-th/0506238}

\bibitem{We03} P.C.~West, {\sl $E_{11}$, $SL(32)$ and central
  charges}, Phys. Lett. B {\bf 575} (2003) 333--342, {\tt
  hep-th/0307098}

\bibitem{DHS}  M.~Dine, P. Huet and N.~Seiberg,
  {\sl Large and small radius in string theory},
  Nucl. Phys. {\bf 322} (1989) 301

\bibitem{DLP} J.~Dai, R.G.~Leigh and J.~Polchinski,
  {\sl New connections between string theories},
  Phys. Lett. {\bf A4} (1989) 2073

\bibitem{AdWLN}
  M.~Abou-Zeid, B.~de Wit, D.~L\"ust and H.~Nicolai,
  {\sl Space-time supersymmetry, IIA/B duality and M-theory},  Phys.\
  Lett.\ B {\bf 466} (1999) 144, {\tt hep-th/9908169}

\bibitem{Ha00}  S.~F.~Hassan, {\sl T-duality, space-time spinors
  and R-R fields in curved backgrounds},  Nucl.\ Phys.\ B {\bf 568}
  (2000) 145, {\tt hep-th/9907152}

%\bibitem{DaNi04} T.~Damour and H.~Nicolai, {\sl Eleven dimensional
%  supergravity and the $E_{10}/$ $K(E_{10})$ $\sigma$-model at low $A_9$
%  levels}, in: Group Theoretical Methods in Physics, Institute of
%  Physics Conference Series No. 185, IoP Publishing, 2005,
%  {\tt hep-th/0410245}

\bibitem{CJS} E.~Cremmer, B.~Julia and J.~Scherk, {\sl
  Supergravity theory in 11 dimensions}, Phys. Lett. B {\bf 76}
  (1978) 409--412

\bibitem{Schwarz:1983wa} J.~H.~Schwarz and P.~C.~West, {\sl Symmetries
  And Transformations Of Chiral N=2 D = 10 Supergravity}, Phys.\
  Lett.\ B {\bf 126} (1983) 301.

\bibitem{Schwarz:1983qr} J.~H.~Schwarz, {\sl Covariant Field Equations
  Of Chiral N=2 D = 10 Supergravity}, Nucl.\ Phys.\ B {\bf 226} (1983)
  269.

\bibitem{Ju85} B.~Julia, in: Lectures in Applied Mathematics, Vol. 21
  (1985), AMS-SIAM, p. 335; preprint LPTENS 80/16

\bibitem{SchnWe04} I.~Schnakenburg and P.~West, {\sl Kac-Moody Symmetries
  of Ten-dimensional Non-maximal Supergravity Theories}, JHEP {\bf
  0405} (2004) 019, {\tt hep-th/0401196}

\bibitem{EnHo04a} F.~Englert and L.~Houart, ``${\cal G}^{+++}$
 invariant formulation of gravity and M-theories: exact BPS
 solutions'', {\it JHEP}, {\bf 0401} (2004) 002, {\tt hep-th/0311255}

\bibitem{KlWe04} A.~Kleinschmidt and P.~West, {\sl Representations
  of ${\cal G}^{+++}$ and the role of space-time}, JHEP {\bf
  0402} (2004) 033, {\tt hep-th/0312247}


\bibitem{Ke04}  A.~Keurentjes, {\sl $E_{11}$: Sign of the times},
  Nucl.\ Phys.\ B {\bf 697} (2004) 302, {\tt hep-th/0402090}

\end{thebibliography}
\end{document}